\documentclass[fleqn,usenatbib]{mnras}

\usepackage{mathptmx}

\usepackage[T1]{fontenc}
\usepackage{ae,aecompl}

\usepackage{graphicx} 
\usepackage{amsmath} 

\usepackage{amssymb} 
\usepackage{amsmath}
\usepackage[utf8]{inputenc}
\usepackage{mathptmx}
\usepackage{breqn}
\usepackage{booktabs}
\usepackage{hyperref}

\hypersetup{urlcolor=blue, colorlinks=true}  

\title[\texttt{CorrSim}: An Observation Simulator]{\texttt{CorrSim}: A Multiwavelength Timing Observation Simulator}

\author[J. A. Paice et al.]{J. A. Paice$^{1,2,3,4}$\thanks{E-mail: \href{mailto:johnapaice@gmail.com}{johnapaice@gmail.com}},
	R. Misra$^{4}$
	P. Gandhi$^{3}$,
	\\
	$^{1}$Centre for Extragalactic Astronomy, Department of Physics, Durham University, South Road, Durham DH1 3LE, UK\\
	$^{2}$Department of Physics and Astronomy, University of Manchester, Oxford Rd, Manchester, M13 9PL, UK\\
	$^{3}$Department of Physics and Astronomy, University of Southampton, Highfield, Southampton, SO17 1BJ, UK\\
	$^{4}$Inter-University Centre for Astronomy and Astrophysics, Pune, Maharashtra 411007, India\\
}

\date{Submitted to MNRAS in original form 2022 AAA XX, Received 2022 AAA XX, Accepted 2022 AAA XX}

\pubyear{2022}

\begin{document}
\label{firstpage}
\pagerange{\pageref{firstpage}--\pageref{lastpage}}
\maketitle

\begin{abstract}
    Studying the rapid variability of many astronomical objects is key to understanding the underlying processes at play. However, a combination of limited telescope availability, viewing constraints, and the unpredictable nature of many sources mean that obtaining data well-suited to this task can be tricky, especially when it comes to simultaneous multiwavelength observations. Researchers can often find themselves tuning observational parameters in real-time, or may realise later that their observation did not achieve their goals.
    Here, we present \texttt{CorrSim}, a program to aid planning of multiwavelength coordinated observations. \texttt{CorrSim} takes a model of a system (i.e. Power Spectra, Coherence, and Lags), and returns a simulated multiwavelength observation, including effects of noise, telescope parameters, and finite sampling. The goals of this are: (i) To simulate a potential observation (to inform decisions about its feasibility); (ii) To investigate how different Fourier models affect a system's variability (e.g. how altering the frequency-dependent lags between bands can affect data products like cross-correlation functions); and (iii) To simulate existing data and investigate its trustworthiness. We outline the methodology behind \texttt{CorrSim}, show how a variety of parameters (e.g. noise sources, observation length, and telescope choice) can affect data, and present examples of the software in action.


	
\end{abstract}

\begin{keywords}
	[ENTER KEYWORDS]
\end{keywords}
	


\section{Introduction} \label{sec:Intro}



When observed at different wavelengths, many astrophysical sources can show startlingly different signals. While these signals are often rapid and complex, they are often all interrelated; they are the result of some base process (or processes) that created them. By studying these different signals, and uncovering the relationships between them, we can thus begin to decode what the underlying processes are.



This general description can be applied to many fields of astrophysics. Simultaneous (or at least quasi-simultaneous) observations on multiple telescopes have been key to the studies of supernovae \citep{Perley_AT2018cow_2019}, Millisecond Pulsars \citep{Draghis_MultibandBW_2019, Papitto_SimultaneousJ1023_2019}, Ultra-luminous X-ray sources \citep{Middleton_SimultaneousMultiwavelengthAstronomy_2017}, Active Galactic Nuclei \citep{McHardy_NGC5548XRayUVLags_2014}, and even Kilonovae \citep{Kasliwal_Kilonova_2017}.

A valuable test case is one particular field: the study of rapid variability in X-ray Binaries. Not only are these complex systems, with a compact object being surrounded by a swirling accretion disc being fed from a companion star, but they are also very dense systems; they are only a few hundred thousand to a few million Schwarzchild radii at the largest length scales. To put this in the terms of time, the dynamical timescale - the timescale at which matter can flow between different parts of the system - is on the order of minutes for typical values \citep{FrankKingRaine_Accretion_2002}. Meanwhile, the systems are only a few tens of lightseconds across, with length scales close to the compact object of around, and below, one lightsecond. This means that two regions - emitting at drastically different wavelengths - can affect one other on timescales of minutes to milliseconds.

Since these sources are so complex and compact, rapid (often sub-second) multiwavelength observations are \textit{key} to understanding how their different regions interact, and what underlying structures are present. Decades of effort here have resulted in a growing collection of correlated observations, revealing complex lags and relations \citep[for just a few examples, see][]{kanbach_correlated_2001, durant_swift_2008, Gandhi_Correlations_2008, gandhi_rapid_2010, Veledina_J1753QPO_2015, gandhi_elevation_2017, Pahari_BWCir_2017, Paice_1820Letter_2019, Vincentelli_J1535_2021}.


However, such studies have had several key problems to overcome. 
For one, a simultaneous observation between two telescopes is difficult to plan, and prone to many issues such as orbital visibility, day/night cycles, packed schedules, and the initial difficulty of obtaining proposals on all required telescopes. And on the night of an observation, low count rates, obfuscating noise (both intrinsic, such as Poisson, and extrinsic, such as read noise), and poorly-chosen time resolutions can all lead to bad or even useless data.

And, even if data are obtained at a high-enough quality, it is only natural to ask: How representative are these data of the source as a whole? A finite observation typically cannot capture the full model of a complex system; this is particularly true for X-ray Binaries, where the behaviour of a source can vary suddenly and dramatically over the course over several weeks \citep[see, e.g.,][]{veledina_swiftj1753ccfs_2017, Paice_J1820Evolution_2021, Thomas_J1820WarpedDisc_2022}, or even minutes \citep{gandhi_furiously_2016, gandhi_elevation_2017}. \citet{Middleton_SimultaneousMultiwavelengthAstronomy_2017} covers many of these problems in greater detail, and notes many fields of study where solving them could lead to significant advances in our understanding (including a case study where simultaneous observations lead to a revolution in the understanding of one source, V404\,Cyg).



To summarise, the main problems are thus; \textit{it is difficult to plan an observation that will deliver data of a high enough quality for us to investigate our questions}, and even if an observation is carried out, \textit{it is not trivial to know if those data correspond well to a model of the source}.

With these issues in mind, we have developed a program, \texttt{CorrSim}. Given both source and observational properties, \texttt{CorrSim} simulates an observation and return various data products, such as lightcurves, cross-correlation functions, and cross-spectral analyses. This will allow for the testing not only of  observational setups, but also various models and how they affect the data produced.

In specific, the motivations of the program are:

\begin{enumerate}
    \item To quantify the minimum data required to perform an analysis; i.e. the minimum length of an observation that would be required, and what magnitude of measurement errors would still produce a reliable result.
    \item To find the conditions under which the shape and significance of obtained cross-correlation functions can be trusted.
    \item To understand the effect of different kinds of measurement errors and incoherent broad band noise on the data analysis.
    \item For smaller data sets and/or those with higher measurement errors, to ascertain which type of analysis would provide the most reliable results.
    \item To investigate the ambiguity of time lags in the analysis, and in the case of periodic signals where the true time lag is greater than one time period away, how to spot this in real data.
\end{enumerate}

We will first detail the workings and assumptions of \texttt{CorrSim} (Sec. \ref{sec:Code}). We will then show the result of putting real, observed source properties into \texttt{CorrSim}, and quantify how factors such as observation length, noise sources, and choice of telescope alter the resultant data (Sec. \ref{sec:Discuss_Params}). We will finally detail two examples of using the program to (i) inform choices about a hypothetical upcoming observation (Sec. \ref{sec:Discuss_Example}) and (ii) test how modified Fourier components affect the cross-correlation functions (Sec. \ref{sec:Discuss_Model}). A glossary of used symbols (Sec. \ref{sec:Glossary}) and description of cross-correlation and Fourier analysis (Appendix \ref{sec:Method}) are also appended at the end of this paper.

We also note the following: X-ray Binaries (XRBs) are referenced throughout this paper as the main motivating factor behind the work, and are used as case studies in our examples. However, \texttt{CorrSim} is built to be a flexible program, and should be useful to any multiwavelength astronomical observations described and investigated the time and Fourier domains.

\section{\texttt{CorrSim}} \label{sec:Code}

We created the program \texttt{CorrSim} as part of this project\footnote{\href{https://gitlab.com/astro_johnapaice/CorrSim}{https://gitlab.com/astro_johnapaice/CorrSim}}. The aim of \texttt{CorrSim} is to present a simulated result of an observation, and then run analysis on that result and compare it to the inputted parameters.

\subsection{Inputs and Outputs}

\begin{figure}
\includegraphics[width=\columnwidth]{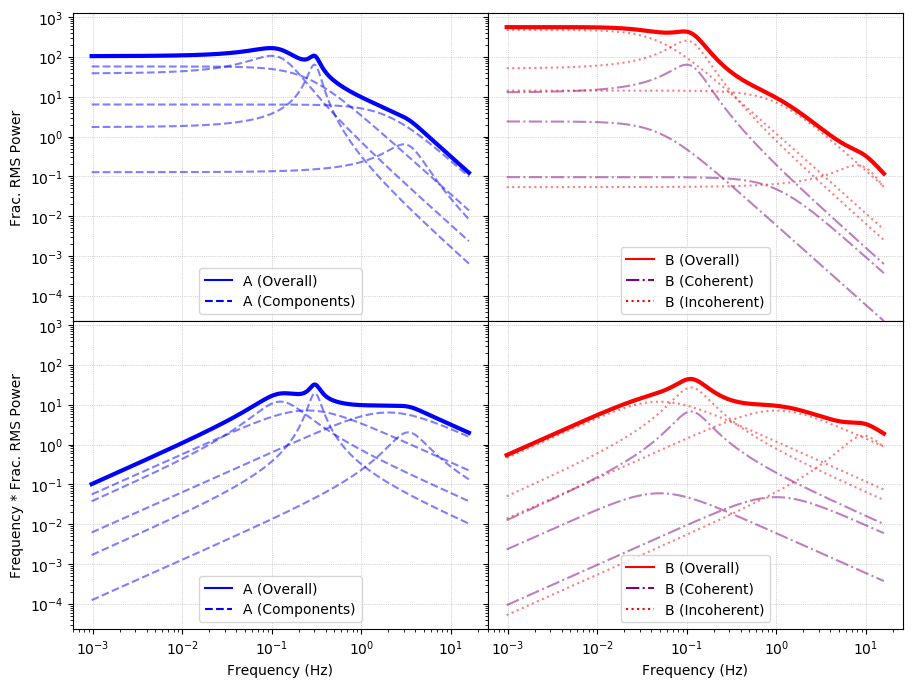}
\caption{Examples of model input power spectra, made up of several Lorentzians. Top row: Fractional RMS power. Bottom row: Fractional RMS power multiplied by frequency. Each component Lorentzian is plotted, as well as the overall summation. Note the coherent and incoherent Lorentzians in Series B; some have the same shape but have different normalisations.}
\label{fig:input_ps_lor}
\end{figure}

\begin{figure}
\includegraphics[width=\columnwidth]{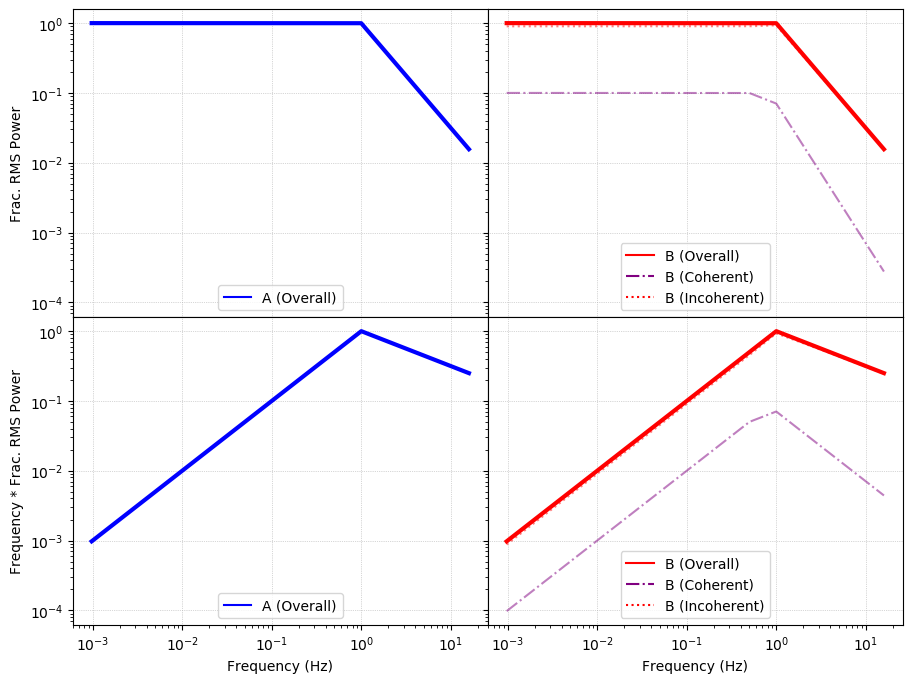}
\caption{Examples of model input power spectra, made up of a broken powerlaw. Top row: Fractional RMS power. Bottom row: Fractional RMS power multiplied by frequency. Note the coherent and incoherent Lorentzians in Series B, and how the coherent power law has a different shape and break frequency.}
\label{fig:input_ps_bpl}
\end{figure}

The inputs here are split into Source (intrinsic to the object) and Observation (intrinsic to the observation) Parameters: 
\begin{itemize}
    \item \textit{Source Parameters}:
    \begin{itemize}
        \item \textbf{Mean count rates} ($\Bar{A}$, $\Bar{B}$) -- Units of counts per second.
        \item \textbf{Fractional RMS values} ($F_{rms, A}$, $F_{rms, B}$) -- units of percent. Defines the variability of the lightcurves. 
        \item \textbf{Model Power Spectra} ($p_{A, Model}$, $p_{B, Model}$). Two power spectral model types are defined in \texttt{CorrSim}: the summation of several Lorentzians (Figure \ref{fig:input_ps_lor}), and a broken power law (Figure \ref{fig:input_ps_bpl}). The shape of the power spectra itself only affects the lightcurve, but the dependence of the coherence on time strongly affects the shape and strength of the correlation function.
        \item \textbf{Model Coherence} ($\gamma_{Model}$). This is defined in \texttt{CorrSim} by the relative coherence of different components in the power spectra.
        \item \textbf{Model Phase/Time lags} ($\delta_{Model}$), as a function of Fourier frequency. Several different ways to define the lags have been provided, relating to the dependence of phase or time with Fourier frequency in either log-log or semi-log space.
        \item \textbf{Red Noise} -- Applies `red' noise, i.e. noise dependant upon frequency. Uses two sub-parameters: `Fractional RMS', which defines the amount of the noise; and `Slope', which defines its dependence on frequency.
        \item \textbf{Poisson Noise} -- Boolean.
        
    \end{itemize}
    \item \textit{Observation Parameters}:
    \begin{itemize}
        \item \textbf{Observation Length} ($T$) -- units of seconds.
        \item \textbf{Time Resolution} ($dT$) -- units of seconds.
        \item \textbf{Scintillation Noise} -- Simulates noise from atmospheric scintillation. Uses several sub-parameters: `Telescope Diameter' (m); `Telescope Altitude' (m); `Exposure Time' (s); `Target Altitude' ($^{\circ}$); `Atmospheric Turbulence Height' (m); and an empirical value \citep{Osborn_Scintillation_2015}.
        \item \textbf{Readout Noise} -- Units of electrons.
    \end{itemize}
\end{itemize}

The outputs are:
\begin{enumerate}
    \item Simulated Lightcurves for each band
    \item Correlation Function Analysis
    \item Fourier Analysis (i.e. Power Spectra, Coherence, and Phase and Time lags as a function of Fourier frequency)
\end{enumerate}

Some of these outputs have sub-parameters for controlling their behaviour, such as segment size and binning for both Correlation Function and Fourier Analysis.

\subsection{Methodology}

\begin{figure}
\includegraphics[width=\columnwidth]{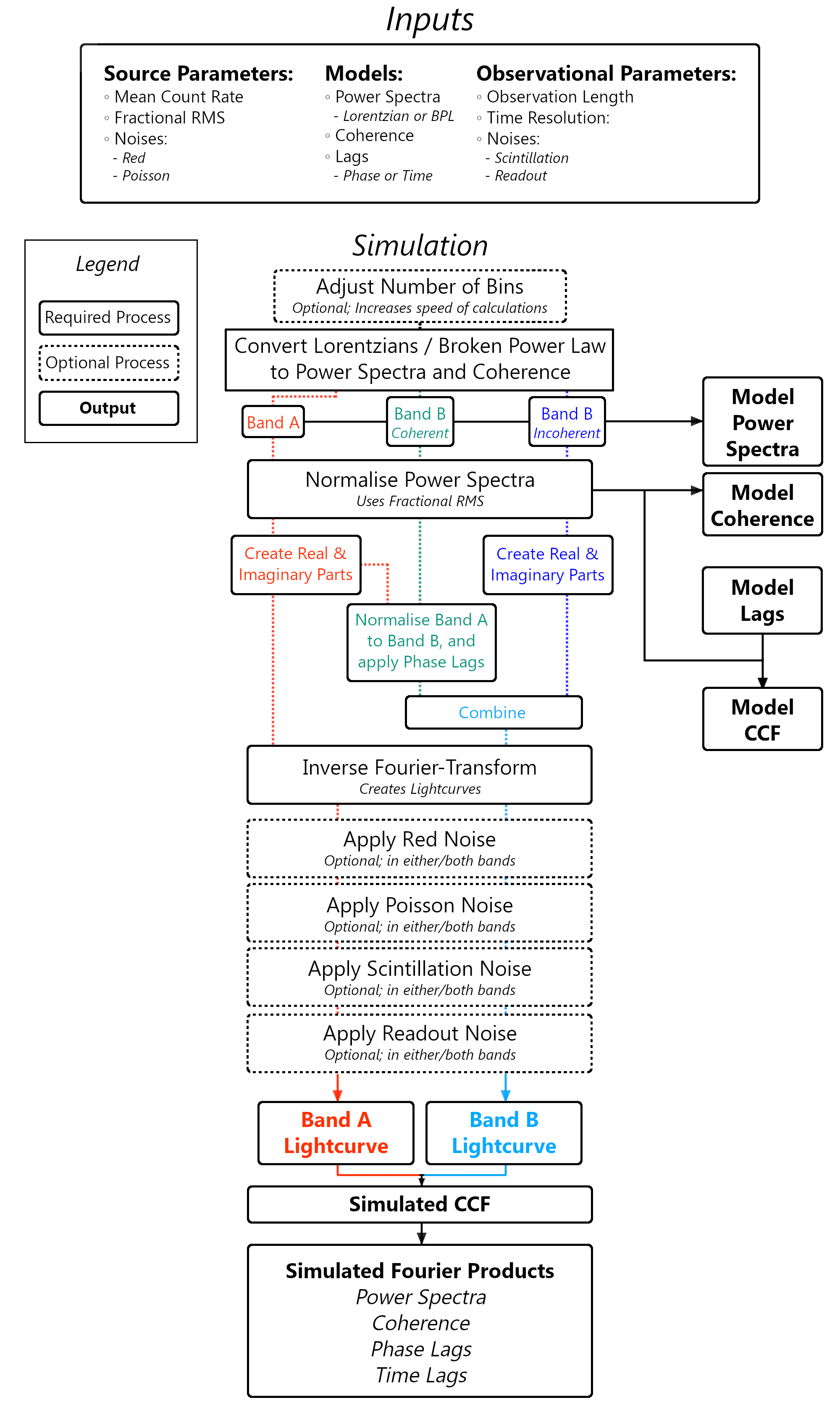}
\caption{Flowchart demonstrating the process that CorrSim uses to create its outputs.}
\label{fig:flowchart}
\end{figure}

A flowchart showing the methodology of \texttt{CorrSim} is shown in Fig. \ref{fig:flowchart}. Below, we go into detail about each of these steps.

Initially, \texttt{CorrSim} (optionally) adjusts the input observation length in order to maximise the speed of the calculations, which go faster with smaller factors. This is done either by adjusting the number of bins to the closest `7-smooth' number (a number with no prime factor larger than 7; \citealt{Berndt_7Smooth_1994}).

\texttt{CorrSim} then creates three power spectra from the details provided -- one for Band A ($p_{A}$), and one for each the coherent and incoherent parts of Band B ($p_{B, Coh}$ and $P_{B, Inc}$ respectively). It then normalises all three to match the fractional RMS values given.

In the next few steps, \texttt{CorrSim} will create the two simulated lightcurves from the power spectra ($S_{A}$ and $S_{B}$), using methodology set out in \citet{TimmerKoenig_PowerLawNoise_1995}. \texttt{CorrSim} first draws two random numbers each for Band A and the incoherent part of Band B, with a standard deviation equal to the respective power spectra. These numbers become the real and imaginary parts of a complex series ($S_{A}$ and $S_{B, inc}$). To make $S_{B, coh}$, the Band A complex series is copied and normalised to Band B, and the phase lags are applied:

\begin{equation}
    S_{B, Coh} = C \times S_{A} \times e^{-i\delta_{Model, Phase}}
\end{equation}

where C = $\sqrt{p_{B,Coh}/p_{A}}$, $i$ is the imaginary number, and $\delta_{Model, Phase}$ are the model phase lags (\textit{not} the time lags). The first bin of $S_{A}$ is adjusted to be equal to the mean count rate multiplied by the number of bins, $\Bar{A}N$ (and $\Bar{B}N$ for $S_{B, coh}$, while the first bin of $S_{B, inc}$ is set to zero). $S_{B, coh}$ and $S_{B, inc}$ are then combined into $S_{B, total}$. Both $S_{A}$ and $S_{B, total}$ are inverse Fourier transformed to turn them into lightcurves, and the fractional RMS is checked against the input values.

Finally, noise is (optionally) added: First, red noise is generated using a similar method from \citet{TimmerKoenig_PowerLawNoise_1995}, and added on to $S_{A}$ and $S_{B, total}$ (in this case, the lightcurves are remade as above). Then, \texttt{CorrSim} adds Poisson noise, calculates and adds scintillation noise, and finally adds readout noise. The result of this is the two final simulated lightcurves for Band A and B.






This is the crux of \texttt{CorrSim}; these lightcurves simulate the data taken of an observed source, i.e. an imperfect representation of the true relationship between signals, and are are analogous to a real observation given the various source and observational parameters input. These lightcurves are not only plotted graphically, but also outputted as comma-separated text files, in case users wish to use them with their own code.

If desired, at this point, \texttt{CorrSim} essentially works in reverse; it takes those lightcurves and runs both Cross-Correlation and Fourier analysis on them, as one would for real, observed data. Much of the Fourier analysis uses functions from \texttt{Stingray} for the calculations. Examples of the plots created are seen in Section \ref{sec:Outputs}.

\subsubsection{Modelling the Correlation Function} \label{sec:ModelCF}

The Correlation Function (CF) is the inverse Fourier transform of the coherence and lags, and thus a model of the function can be reproduced purely from these properties. This is based on the mathematical definitions of coherence presented in \citet{Vaughan_Nowak_1997}, and is defined in Section \ref{sec:Method_CCF}. In \texttt{CorrSim}, a model of the Correlation Function is produced from the initial coherence and lag inputs, and the methodology is described here.

A Fourier transform creates a complex series, which can be wholly described by its amplitude and its arguments. The amplitude of this series is defined as:

\begin{equation}
    U = \gamma_{Model} * \sqrt{(p_A)({F_{rms}}_A) * (p_B)({F_{rms}}_B))}
\end{equation}

where $U$ is the amplitude, $\gamma_{Model}$ is the model Coherence, and $p$ and $F_{rms}$ are the power spectra and fractional RMS values (for each band A and B). The arguments are then taken to be the phase lags. This complex series is then inverse Fourier transformed, and the real part is taken. The result is normalised with a constant ($V$) thusly:

\begin{equation}
    V = \frac{1}{2 * dT * \sqrt{{F_{rms}}_A * {F_{rms}}_B}}
\end{equation}

where $dT$ is the time resolution of the data. After the normalisation is applied, the final product is the model Correlation Function.

\subsubsection{Converting Phase to Time} \label{sec:CorrSim_PhaseToTime}

Phase lags, when combined with the frequency they are found at, can be converted to time lags using Equation \ref{eqn:timelags}.




However, there is a limitation to phase lags; \texttt{CorrSim} adopts the convention that phase lags are given between $\pm\pi$. If the actual phase lags are outside of this range, say between $\pi$ and $2\pi$\,radians, then \texttt{CorrSim} will show that they are between $-\pi$ and 0\,radians due to the cyclical nature of sine waves. This limitation passes on to time lags; a `true' time lag that is between $+\pi$ and $+2\pi$\,radians in phase will be represented as a negative time lag.

There is a solution -- or, at least, a mitigation -- to this. By using the correlation function, we can see where sources of correlation (or anti-correlation) are occurring and find which frequency bins are most likely to be correct. Then, assuming that phase lags follow a reasonably continuous distribution, we shift them by $\pm2\pi$ radians to minimise discontinuities.

This method is intrinsic to \texttt{CorrSim}. A `Reference Frequency' parameter is set (default 1\,Hz) which is assumed to be correct. Then, \texttt{CorrSim} looks at the next lowest frequency. That point is compared to the \textit{mean of the previous three points}. If the difference is greater than $\pi$, it \textit{and all points at lower frequencies than this one} are shifted by $\pm2\pi$ to minimise the discontinuity. This is repeated for every frequency bin. The same procedure is then conducted, but instead going to higher frequencies. These corrected phase lags\footnote{The corrected phase lags themselves are not plotted; the phase lags shown are the ones gained from the Fourier analysis, without any shifting.} are then converted to time lags using the above equation.

This is a method that has its drawbacks -- any processes constrained to just a single frequency bin, such as Quasi-Periodic Oscillations, may be more than $\pi$ away from phase lags in adjacent bins, and this will thus be misrepresented in the time lags. Close inspection of the time lag plots, and comparison with the CF, may be required to investigate this possibility.

\subsection{Assumptions, Models and Noise}

Several assumptions have been made in producing \texttt{CorrSim} -- these have been for simplicity and ease of use, and none should strongly affect the primary results. They are detailed here.

\begin{enumerate}
    \item \textbf{Power Spectra}
    
    The broken power law, and its handling of coherence, is a simplification of the behaviour of a source. Lorentzians, while also being limited in their own ways, offer much finer control.
    
    \item \textbf{Phase and Time Lags}
    
    \texttt{CorrSim} approximates the phase and time lags using a series of constant, linear, power law, or polynomial distributions. This does not trivially handle the true lag behaviour -- which may be better represented by other distributions -- but the versatility of \texttt{CorrSim} in being able to handle any number of segments mitigates this simplification. In calculating the time lags, \texttt{CorrSim} also assumes that phase lags are roughly continuous and do not deviate by more than $\pi$ radians between frequency bins.
    
    \item \textbf{Simultaneous, Continuous Lightcurves}
    
    \texttt{CorrSim} creates lightcurves that are simultaneous, sampled at the same rate, and have no gaps, and assumes this is the case during its analysis. While this is not wholly realistic for most multiwavelength observations, it serves as a good approximation, and the simultaneous sampling vastly speeds up the calculation of the correlation function.
    
    If a user wishes to analyse a lightcurve that is affected by gaps, then they may take the lightcurve data that \texttt{CorrSim} generates and create the gaps manually, and then use their own code to do the analysis.
    
    \item \textbf{Noise}
    
    \texttt{CorrSim} currently allows for four sources of noise. The way this noise is calculated is a simplification of processes within the source and the instrument. Their methods are detailed here.
    \begin{enumerate}
    \item \textit{Red noise}. This is noise that is more significant at higher powers. Using methods set out in \citet{TimmerKoenig_PowerLawNoise_1995}, this noise is modelled by drawing two random Gaussian numbers and then multiplying them by a value proportional to the model power spectra for each frequency bin. These two resultant numbers then become the real and imaginary parts of the complex array which will later create the lightcurve.
    
        \item \textit{Poisson noise}. This noise accounts for random variation in the source. This is modelled by taking the lightcurve, and for each bin, drawing a random number from a Poisson distribution ($\lambda$ = counts in that bin).
        
        \item \textit{Scintillation noise}. This noise arises from scintillation seen by ground-based telescopes, caused by variations in the atmosphere. \texttt{CorrSim} uses a modified version of Young's approximation, presented in Equation 7 of \citet{Osborn_Scintillation_2015}, which is observation- and telescope-specific; the required parameters include the telescope's diameter and altitude, the distance of the source from zenith, the height of atmospheric turbulence, exposure time, and an empirical, telescope-specific constant. The equation gives the variance of the scintillation noise. Therefore, this noise is modelled by taking the lightcurve, and for each bin, drawing a random number from a Gaussian distribution, with the mean equal to the counts in that bin, and the standard deviation equal to the square root of Young's approximation.
                
        \item \textit{Readout noise}. This noise arises from the imperfect amplification of a signal in the CCD, where the true charge on a number of electrons is misread. This noise is modelled by drawing from a Gaussian distribution centred on zero, with the standard deviation equal to the readout noise in electrons, for each bin, and then adding that number on to the value of the bin.
        
    \end{enumerate}
\end{enumerate}
	
\subsection{Outputs} \label{sec:Outputs}

\begin{figure*}
\includegraphics[width=\textwidth]{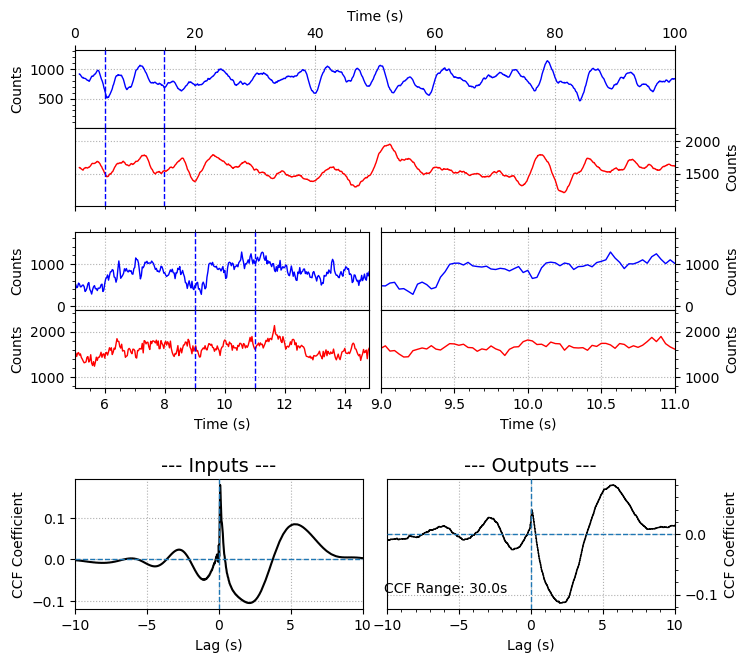}
\caption{The first half of a compilation plot provided by \texttt{CorrSim}. \textbf{Top, Middle:} 100\,s, 10\,s, and 2\,s segments of the produced lightcurves (Band A, blue; band B, red). The 100\,s segment has a moving average function applied over fifty points. The dashed lines represent the ranges of the insets. \textbf{Bottom:} Model (left) and simulated (right) correlation functions (Band B versus Band A, i.e. a peak at positive lags shows Band B lagging Band A), with the latter also showing the range over which it was averaged. The model correlation function was calculated from the input Coherence, Power Spectra, and Lags (see Figure \ref{fig:compilationfourier}, while the reproduced correlation function was produced from the lightcurves.}
\label{fig:compilationlcs}
\end{figure*}

\begin{figure*}
\includegraphics[width=\textwidth]{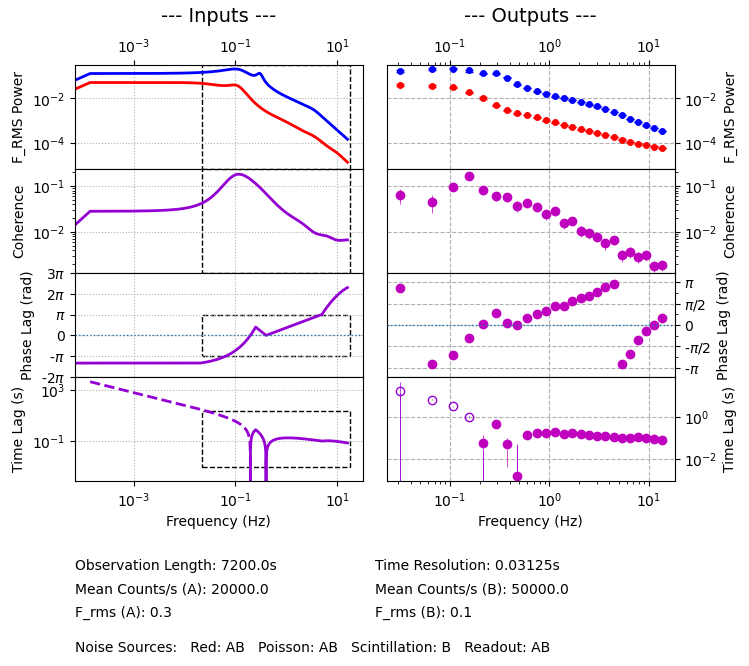}
\caption{The second half of a compilation plot provided by \texttt{CorrSim} -- input Fourier properties (left) and results from Fourier analysis of the lightcurves (right). Since the outputs are calculated by splitting up the lightcurve and averaging over several segments, they only cover a smaller range; black dashed boxes on the left represent this range. \textbf{First Row:} Power Spectra of Band A (blue) and B (red). \texttt{CorrSim} has attempted to remove Poisson noise in the bands. \textbf{Second Row:} The coherence between the bands. Note how the presence of uncorrelated noise reduces the input coherence. \textbf{Third Row:} The phase lags.  For the output, since phase lags outside of $\pm\pi$ are meaningless, the plot wraps around at these ranges. \textbf{Fourth Row:} The time lags. Negative time lags are represented as a dashed line for the input, and open circles for the output. In \texttt{CorrSim}, a reference point can given for calculating the time lags from the phase lags, i.e. stating at which frequency the phase lags can be trusted to be the true value, and not multiples of 2$\pi$ prior or hence. \textbf{Bottom Information:} Key input values. For the noise sources, the letters (A, B, AB, or n) represent which bands that noise is present in, if any.}
\label{fig:compilationfourier}
\end{figure*}

Figure \ref{fig:compilationlcs} shows example produced lightcurves and both the input and output correlation functions. Figure \ref{fig:compilationfourier} shows examples of both inputs and outputs of \texttt{CorrSim}, and some observation parameters.

The Fourier properties of this example are based on data from the X-ray Binary MAXI\,J1820+070. The input parameters, such as the $F_{rms}$, power spectra, and phase lags, are all approximations to those found in the analysis carried out by \citet{Paice_J1820Evolution_2021}, during epoch 4 (its initial rising hard state, 25\,days after peak X-ray brightness, 37\,days after the outburst was first detected). Meanwhile, the noise parameters have been selected to correspond with an Optical (Band A) and X-ray (Band B) observation (taken by the HiPERCAM instrument at the Gran Telescopio Canarias, Roque de Los Muchachos, La Palma and the NICER instrument on the ISS respectively).

\section{Discussion I: Comparison of Parameters} \label{sec:Discuss_Params}

\begin{figure*}
\includegraphics[width=\textwidth]{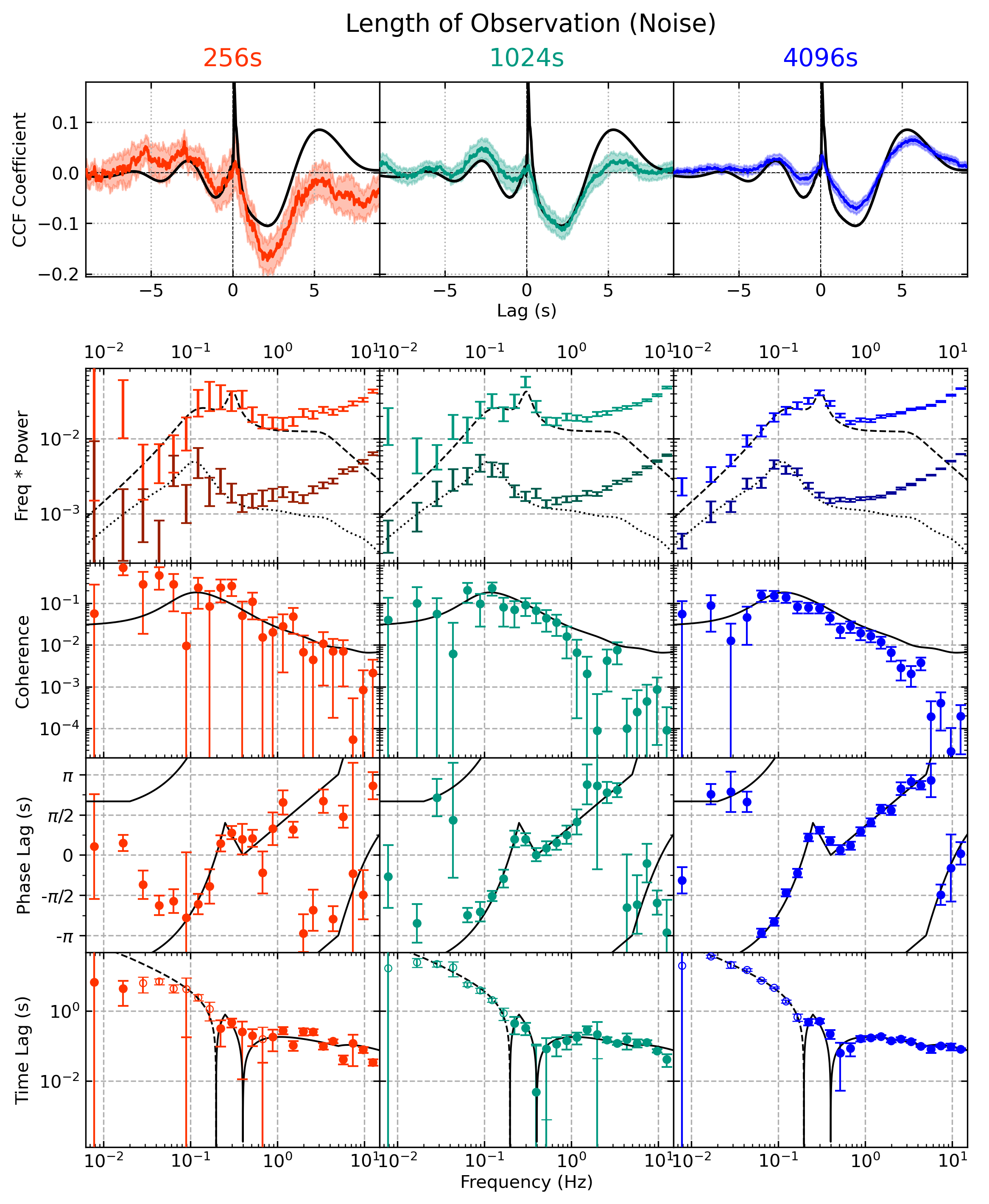}
\caption{Three different runs of \texttt{CorrSim} with different lengths of observations; 256\,s, 1024\,s, and 4096\,s (with a time resolution of $2^{-5}$ (0.004)\,s, this is equivalent to 2$^{13}$, 2$^{15}$, and 2$^{17}$\,bins respectively) with default noise added. Note the significant deviation at higher frequencies, especially with shorter observations. The correlation function and coherence are also lower, even in the longest observations -- the model correlation function and coherence do not predict the deviation due to noise.}
\label{fig:CorrSimCompare_ObsLengthNoise}
\end{figure*}

\begin{figure*}
\includegraphics[width=\textwidth]{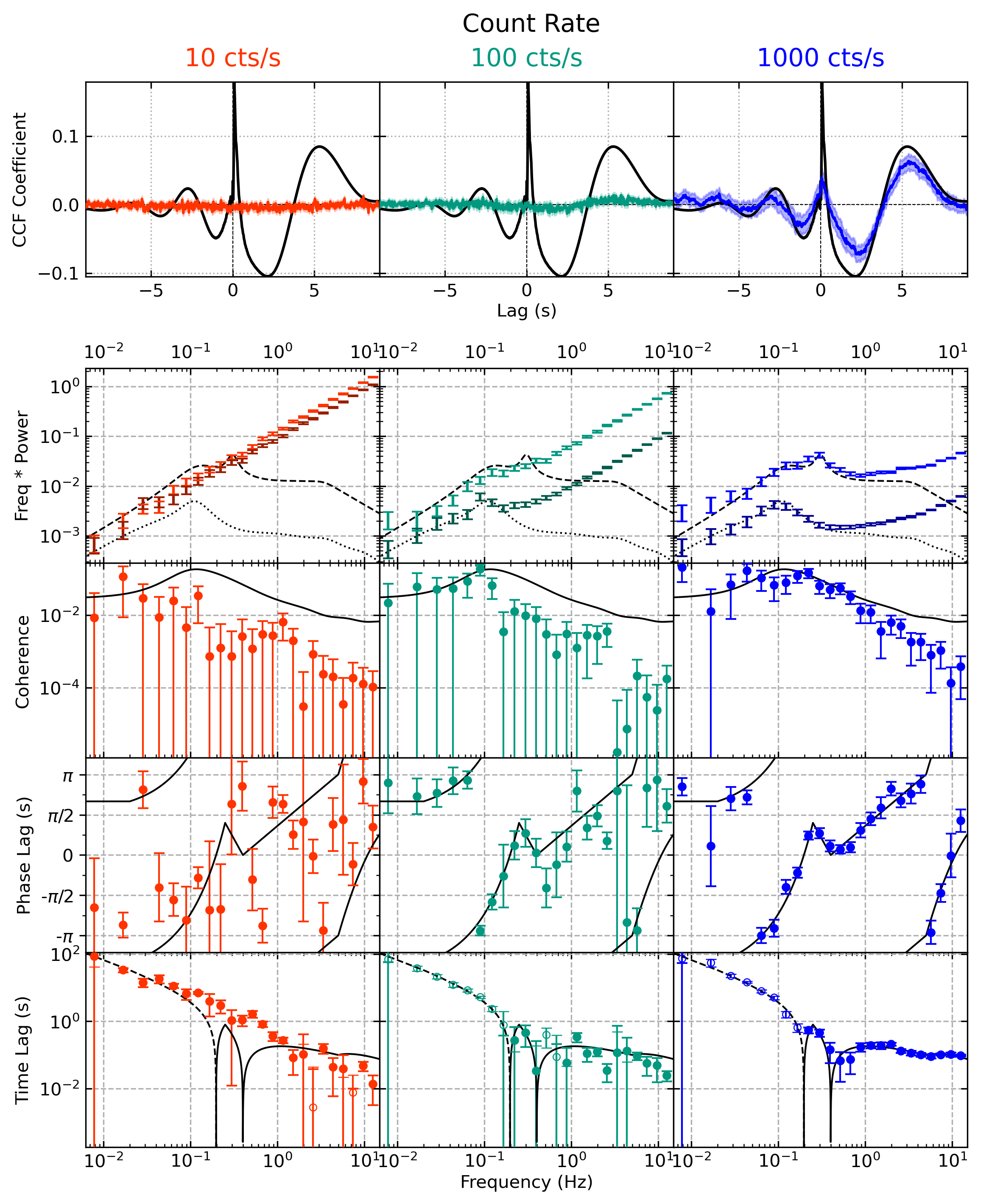}
\caption{Three different runs of the \texttt{CorrSim} code with different count rates; 10\,cts/s, 100\,cts/s, and 1000\,cts/s for Band A (and 50, 500, and 5000\,cts/s for Band B respectively). Note how the lower count rates have significantly lower coherence and magnitudes of the CF. Also see how power spectra can be recovered much more easily than other Fourier products.}
\label{fig:CorrSimCompare_CountRate}
\end{figure*}

\begin{figure*}
\includegraphics[width=\textwidth]{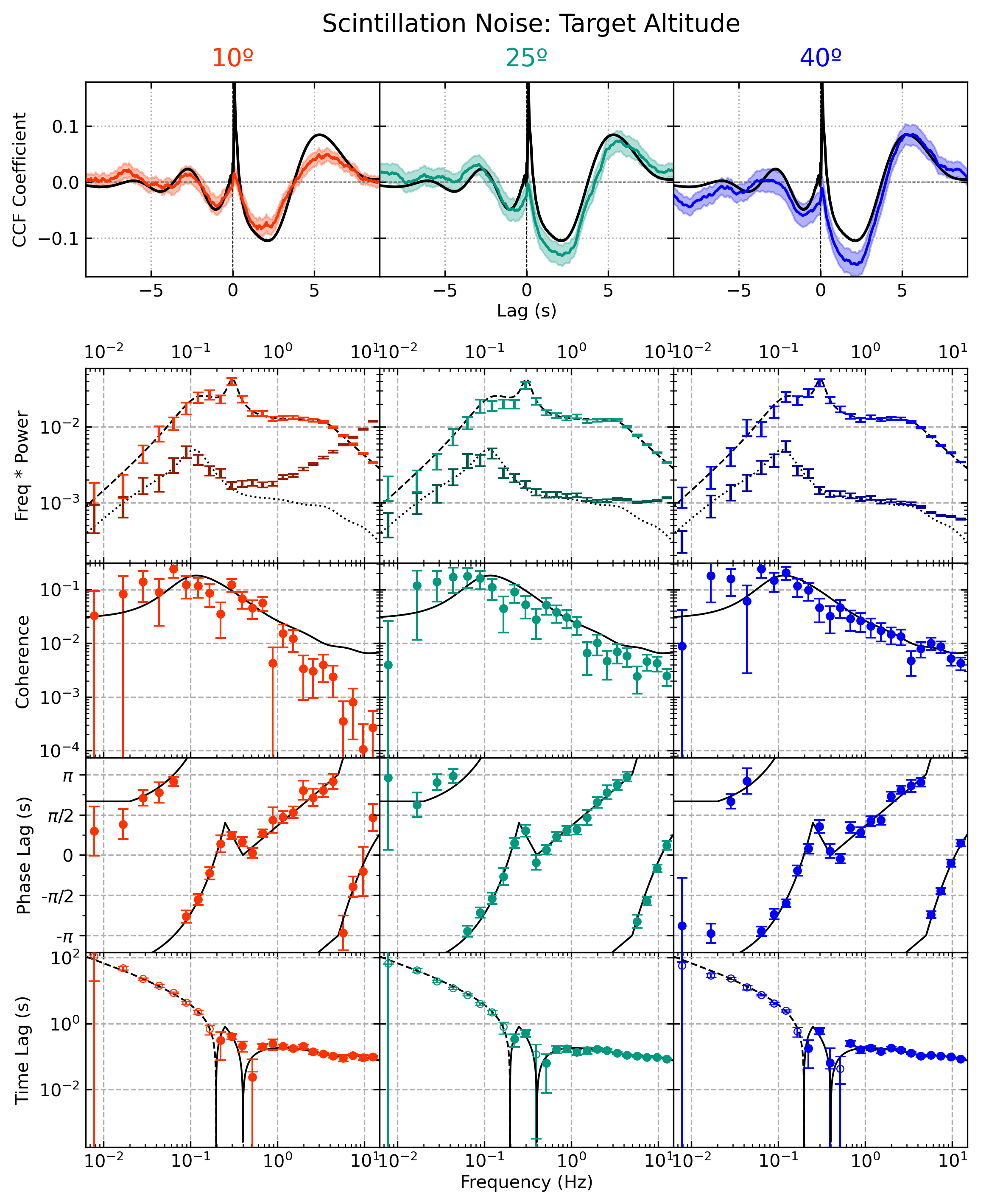}
\caption{Three runs of \texttt{CorrSim} with different target altitudes and its effect on scintillation noise; 10$^{\circ}$, 25$^{\circ}$, and 40$^{\circ}$. All other noise sources were turned off for these simulations, and scintillation noise was only applied to Band B (to simulate a Space + Ground based observation). All other scintillation noise values were kept at defaults. Note how scintillation noise strongly affects a target at 10$^{\circ}$, yet has minimal effect on a target at 40$^{\circ}$ (and thus higher altitudes).}
\label{fig:CorrSimCompare_Scin_Altitude}
\end{figure*}

These simulations allow us to compare different aspects of observations, and see their effects on the final products with respect to the initial models. In this section, we will detail how several different parameters can affect the final products. There are many parameters that can be varied with CorrSim, and not all can be shown here; more figures can be seen in the Supplementary Material, found online. As an example, Figures \ref{fig:CorrSimCompare_ObsLengthNoise}--\ref{fig:CorrSimCompare_Scin_Altitude} demonstrate the effects of varying the observation length, the count rate, and the target altitude.

Each figure shows three different setups and details their effect on the simulated correlation function and Fourier outputs. The models are given by black lines, and the outputs by coloured lines.

The first row is the Cross-Correlation Function (CCF), a particular kind of Correlation Function that functions with equally-spaced bins in a gapless lightcurve, with the shaded region indicating the standard error; note that the model CF does not take effects of noise into account, which reduces coherence and thus the correlation coefficient; the model CF therefore may not approximate a CF created from a theoretically infinite observation, and would instead have a greater magnitude.

The second row shows the power spectra. The black dashed lines are the model Band A power spectra, with lighter coloured uncertainties representing their outputs, and the black dotted lines are the model Band B power spectra with the darker coloured uncertainties representing their outputs. Like the CF, the model power spectra do not take the effects of noise into account.

The third row shows the coherence. These models, like the CF and power spectra, do not take the effects of noise into account.

The fourth row shows the phase lags. This is constrained between $\pm\pi$, and additional models at normalisations of $\pm2\pi$ are also shown; the `middle' model, dominant at 1\,Hz, is the `correct' one. Model phase lags are not affected by noise sources in a predicatable way.

The fifth row shows the time lags. Solid black lines show the model time lags, while dashed black lines show the inverse model time lags, for representing negative lags. Similarly, the solid coloured circles show the outputs, and the open coloured circles show their inverse. Time lags are fixed at a defined reference frequency, and then shifted depending on what causes the smallest discontinuities, as described in Section \ref{sec:CorrSim_PhaseToTime}. As they are calculated from the phase lags, the model time lags are also not affected by noise sources in a predicatable way.

Tables \ref{tab:CorrSimCompare_Defaults}--\ref{tab:CorrSimPhaseLags} show the default inputs for these tests (i.e. the values that will be used unless specified otherwise); these approximate a reasonably bright source observed with a reasonably sensitive telescope. The Lorentzian and Phase Lag distributions in Tables \ref{tab:CorrSimLorentzians} \& \ref{tab:CorrSimPhaseLags} are approximations to the real Fourier properties found in the X-ray Binary MAXI\,J1820+070 during its April 2018 hard state (see Sec. \ref{sec:Outputs}).


\begin{table*}
	\centering
    \caption{Default values for the comparison data.}
    \begin{tabular}{c c c c}
    \toprule 
    \textbf{Group} & \textbf{Parameter} & \textbf{Default Value} & \textbf{Notes}\\
    \midrule
    Observation & Length of Observation (s) &  2048  & \\
     & Time Resolution (s) &  0.03125  & $=2^{-5}$ \vspace{0.12cm}\\
    Band A & Mean Count Rate (cts/s) &  1000  & \\
     & Fractional RMS (\%) &  0.31  & \\
     & Red Noise? (Y/N) & Y & See Footnote 1 \\
     & Poisson Noise? (Y/N) & Y & \\
     & Readout Noise? (Y/N) & Y & 4.5 electrons \\
     & Scintillation Noise? (Y/N) & N & \vspace{0.12cm}\\
    Band B & Mean Count Rate (cts/s) &  5000  & \\
     & Fractional RMS (\%) &  0.11  & \\
     & Red Noise? (Y/N) & Y & See Footnote 1 \\
     & Poisson Noise? (Y/N) & Y & \\
     & Readout Noise? (Y/N) & Y & 4.5 electrons \\
     & Scintillation Noise? (Y/N) & Y & See Footnote 2 \vspace{0.12cm}\\
    Fourier & Power Spectra Model & Lorenztians & See Table \ref{tab:CorrSimLorentzians} \\
     & Lag Model & Phase & See Table \ref{tab:CorrSimPhaseLags} \vspace{0.12cm}\\
    Plotting & CF Range (s) & 30 & \\
     & CF Binning & 0 & \\
     & Fourier Segment Length (Bins) & $2^{12}$ & \\
     & Fourier Rebinning Factor & 1.3 \\
     & Reference Frequency (Hz) & 1 & See Footnote 3\\

    \bottomrule
    \end{tabular}
    \vspace{0.2cm}
    
     \footnotesize{\textit{1} For Band A and B, the Fractional RMS of the red noise is 0.2 and 0.03 respectively, analogous to red noise from X-rays ans Optical. The slope of the red noise is -2 for both.}\\
     \footnotesize{\textit{2} Using reasonable values, analogous to the New Technology Telescope (NTT) at La Silla, Chile \citep{TarenghiWilson_NTT_1989}:\\
     Telescope Diameter = 3.58\,m; Telescope Altitude = 2400\,m; Exposure Time = Time Resolution-1.5\,ms (i.e. `Deadtime' of 1.5\,ms); Target Altitude = 40$^\circ$; Turbulence Height = 8000; Empirical Coefficient C$_{Y}$ = 1.5}
     \\
     \footnotesize{\textit{3} Frequency at which the phase lag is be assumed to be correct (i.e. not shifted by $\pm2\pi$)} \\

	\label{tab:CorrSimCompare_Defaults}
\end{table*}

    



\begin{table*}
	\centering
    \caption{Lorentzians parameters}
    \begin{tabular}{c c c | c c c c }
    
    \toprule 
    \multicolumn{3}{c|}{\textbf{Band A}} & \multicolumn{4}{c}{\textbf{Band B}} \\
    \cmidrule(lr){1-3}
    \cmidrule(lr){4-7} 
    \textbf{Norm} & \textbf{Width} & \textbf{Midpoint} &
    \textbf{Norm} & \textbf{Width} & \textbf{Midpoint} & \textbf{Coherence Fraction}\\
    \midrule

    45  & 0.5   & 0     & 75 & 0.1 & 0   & 1/200 \\
    40  & 4     & 0     & 50 & 0.1 & 0.1 & 1/5 \\
    25  & 0.15  & 0.1   & 45 & 2   & 0   & 1/150 \\
    10  & 0.1   & 0.3   & 6  & 4   & 3   & 0 \\
    3   & 3     & 3     & 3  & 10  & 8   & 0 \\
        &       &       & 1  & 20  & 25  & 0 \\

    \bottomrule
    \end{tabular}
	\label{tab:CorrSimLorentzians}
\end{table*}

\begin{table*}
	\centering
    \caption{Phase Lag Parameters$^{1}$.}
    \begin{tabular}{c c c c c c c c c c c c c c c c c c c }
    
    \toprule 
    \textbf{Distribution} & \textbf{Freq. 1$^{2}$} & \textbf{Lag 1$^{3}$} & \textbf{Freq. 2$^{2}$} & \textbf{Lag 2$^{3}$} & \textbf{Freq. 3$^{2}$} & \textbf{Lag 3$^{3}$}\\
    \midrule

    
    Constant (Phase)& 0.001 & -4$\pi$/3 & 0.02  & -         & -         & - \\
    Power           & 0.02  & -4$\pi$/3 & 0.25  & 2$\pi$/5  & -         & - \\
    Linear          & 0.25  & 2$\pi$/5  & 0.4   & 0         & -         & - \\
    Linear          & 0.4   & 0         & 5     & $\pi$     & -         & - \\
    Polynomial      & 5     & $\pi$     & 200   & $\pi$     & 28        & 5$\pi$/2 \\

    \bottomrule
    \end{tabular}
    \vspace{0.2cm}
    
    \footnotesize{\textit{1} Outside of the specified frequencies, the lag is set to 3$\pi$/4}\\
    \footnotesize{\textit{2} Units of Hz}\\
    \footnotesize{\textit{3} Units of Radians} \\
	\label{tab:CorrSimPhaseLags}
\end{table*}



\subsection{Error Dependency}

For the following tests, all parameters were kept as those in Tables \ref{tab:CorrSimCompare_Defaults}--\ref{tab:CorrSimPhaseLags}, unless otherwise stated. Note that these data are empirical, obtained from a single run of \texttt{CorrSim} each time, rather than model values.

\subsubsection{Error in the Correlation Function}

Another way to illustrate how different parameters affect the observation is by looking at the standard error in the CCF. The assumption here is that if the error is lower, the simulated CCF will be closer to the true model shape.

\begin{figure}
\includegraphics[width=\columnwidth]{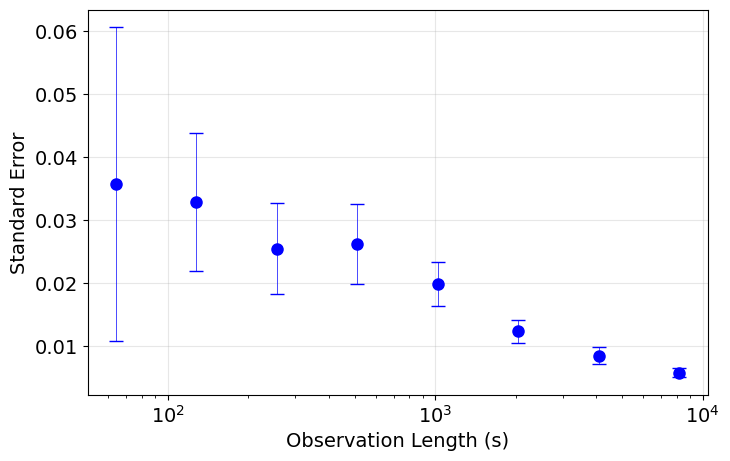}
\caption{Empirical dependency of the standard error in the CCF -- and its standard deviation -- on observation length.}
\label{fig:CorrSim_StandardError}
\end{figure}

We ran \texttt{CorrSim} again with eight different lengths of observation (64, 128, 256, 512, 1024, 2048, 4096, and 8192 seconds). A CCF was constructed from 30\,s segments, and standard error was calculated for each bin from those segments, as usual. For this demonstration, the middle third of the CCF was selected (from -10\,s to +10\,s lags), and then the mean and the standard deviation of the standard error were calculated. The results are plotted in Figure \ref{fig:CorrSim_StandardError}.

\subsubsection{Percentage Error in the Fourier Components}

We can also look at the magnitude of the errors. For this, we assume that smaller errors mean that the observation is closer to the model, which we consider a reasonable assumption for illustrative purposes. By varying a parameter, we can thus see how much it affects the accuracy of the observation by quantifying the change in magnitude of the error (relative to the value of the bin).

\begin{figure}
\includegraphics[width=\columnwidth]{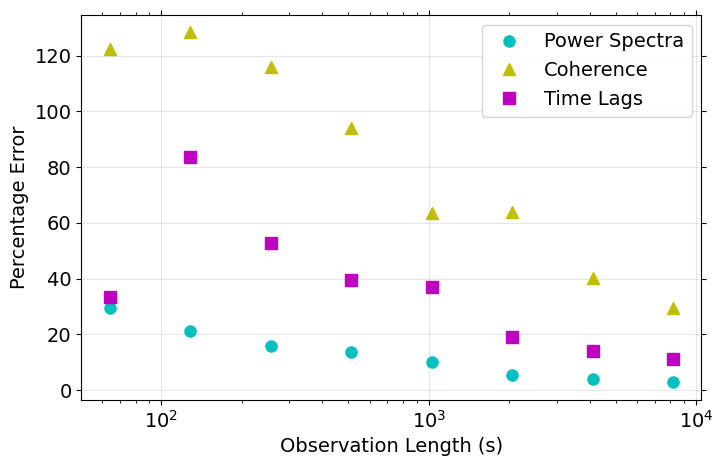}
\caption{Empirical dependency of the median absolute percentage error on observation length.}
\label{fig:CorrSim_FourierError}
\end{figure}

Similar to the previous section, we ran \texttt{CorrSim} again with eight different lengths of observation (64, 128, 256, 512, 1024, 2048, 4096, and 8192 seconds), while the Fourier Segment Length was set to 2$^{10}$ bins. Fourier plots were made for each run. Then, for each of the Power Spectra, Coherence, and Time Lags, the median absolute percentage error was calculated over all bins. The results are plotted in Figure \ref{fig:CorrSim_FourierError}.

\subsubsection{Detecting a Sub-Second Lag}

We can also vary how correlated the lightcurves are, and find how that affects the resultant CCF. For this, we altered the Band B Lorentzians: all coherence fractions were set to 0, and then the second Lorentzian was set to a normalisation of 10 and a midpoint of 1. Its coherence fraction was then varied, and the resultant CCF was plotted in each instance.

\begin{figure}
\includegraphics[width=\columnwidth]{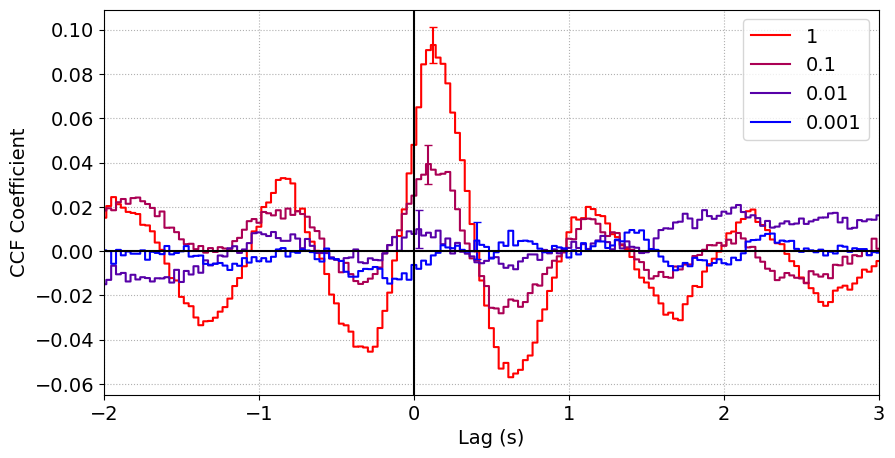}
\caption{Four CCFs created by varying the coherence fraction of a single Lorentzian at 1 Hz, with all other Lorentzians at 0 coherence. Note how the magnitude of the sub-second lag changes with each coherence fraction. Representative errors are plotted.}
\label{fig:CorrSim_CCFLagTest}
\end{figure}

Figure \ref{fig:CorrSim_CCFLagTest} shows a selection of the results. As expected, while the sub-second lag is clearly identifiable at a coherence fraction of 1, it is almost completely gone at a fraction of 0.001. Under these source and observational conditions, this graph can thus tell us that a coherence fraction of 0.01 is the minimum to detect a sub-second lag, while values of at least 0.1 are much preferred.

\section{Discussion II: Example Usage} \label{sec:Discuss_Example}

We will now demonstrate how an astronomer may use \texttt{CorrSim} to plan and maximise an observation. Let us say they want to observe a source similar to that given in Tables \ref{tab:CorrSimCompare_Defaults} \& \ref{tab:CorrSimLorentzians}, but with a lower count rate (100 and 1000\,counts/s in bands A and B respectively), and they have 1024\,s of observation time with a time resolution of 0.03\,s (2$^{-5}$). The source has been known to show a broad precognition dip and a sub-second lag, as well as a small feature in the phase lags (described in Table \ref{tab:CorrSimExample_PhaseLags}). It is the latter two in particular -- the sub-second lag and the feature in the phase lags -- that the observer wants to study.

\begin{table*}
	\centering
    \caption{Example -- Phase Lag Parameters$^{1}$.}
    \begin{tabular}{c c c c c c c}
    
    \toprule 
    \textbf{Distribution} & \textbf{Freq. 1$^{2}$} & \textbf{Lag 1$^{3}$} & \textbf{Freq. 2$^{2}$} & \textbf{Lag 2$^{3}$} & \textbf{Freq 3$^{2}$} & \textbf{Lag 3$^{3}$}\\
    \midrule

    
    Linear      & 0.1   & 3$\pi$/4  & 0.25  & $\pi$/5   & - & - \\
    Linear      & 0.25  & $\pi$/5   & 0.3   & 2$\pi$/3  & - & - \\
    Linear      & 0.3   & 2$\pi$/3  & 0.4   & 0         & - & - \\
    Linear      & 0.4   & 0         & 5     & $\pi$     & - & - \\
    Polynomial  & 5     & $\pi$     & 200   & $\pi$     & 28 & 5$\pi$/2 \\

    \bottomrule
    \end{tabular}
    \vspace{0.2cm}
    
    \footnotesize{\textit{1} Outside of the specified frequencies, the lag is set to 3$\pi$/4}\\
    \footnotesize{\textit{2} Units of Hz}\\
    \footnotesize{\textit{3} Units of Radians} \\
	\label{tab:CorrSimExample_PhaseLags}
\end{table*}

\begin{figure*}
\includegraphics[width=\textwidth]{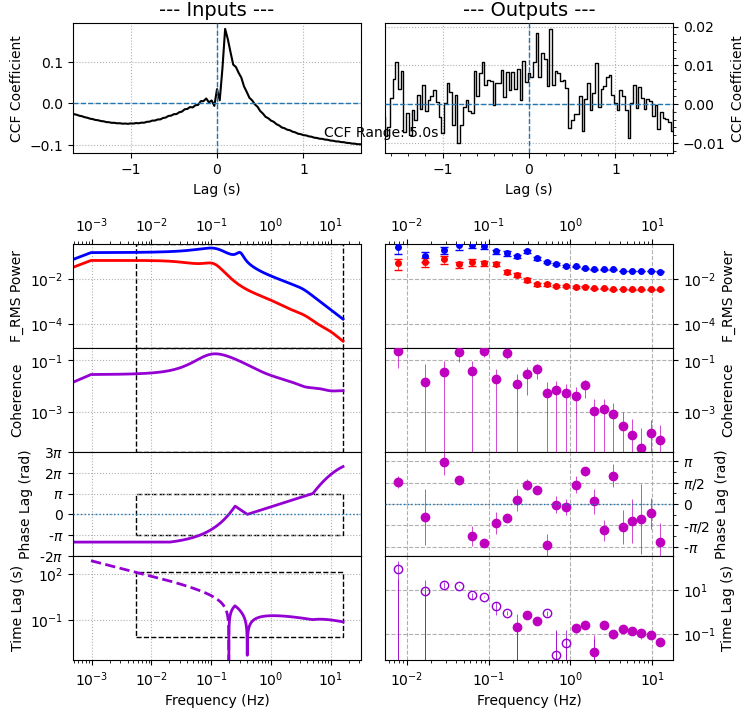}
\caption{Fourier inputs and outputs for the example observation. Note the lack of clarity in the outputs, especially in the phase lags, compared to the inputs.}
\label{fig:CorrSimExample_Output}
\end{figure*}

By running \texttt{CorrSim} with these parameters, our astronomer gets what is shown in Figure \ref{fig:CorrSimExample_Output}. Initially, they are not pleased with the results -- the signal-to-noise ratio (S/N) is too low for their purposes, with the feature in the phase lags indistinguishable from random variations, and the sub-second lag not as clear nor as smooth as hoped. Now they have seen that their plan would not give them what they want, they make a decision to change the setup of the observation.

What can our astronomer do? They have a selection of possibilities in front of them, but ordinarily, it might mainly involve guesswork (and luck) to decide between them. From their options, they want to investigate three possibilities:

\begin{itemize}
    \item \textbf{1: Double the observation length.} By increasing the length of the observation, one would increase the S/N.
    
    \item \textbf{2: Decrease time resolution by a factor of four.} Smaller time bins can cause problems like readout noise to become more significant, or missing some photons due to deadtime. Increasing the bin size, and thus lowering the time resolution, would reduce such effects.
    
    \item \textbf{3: Using a better telescope.} Fortunately, our intrepid astronomer knows of another telescope they can use for Band A. This telescope has five times the sensitivity (i.e. the mean count rate in Band A would become 500) but they can only use it for half the time (i.e. observation length would be 512s). Is there a net benefit to using this telescope?
\end{itemize}

\begin{figure*}
\includegraphics[width=\textwidth]{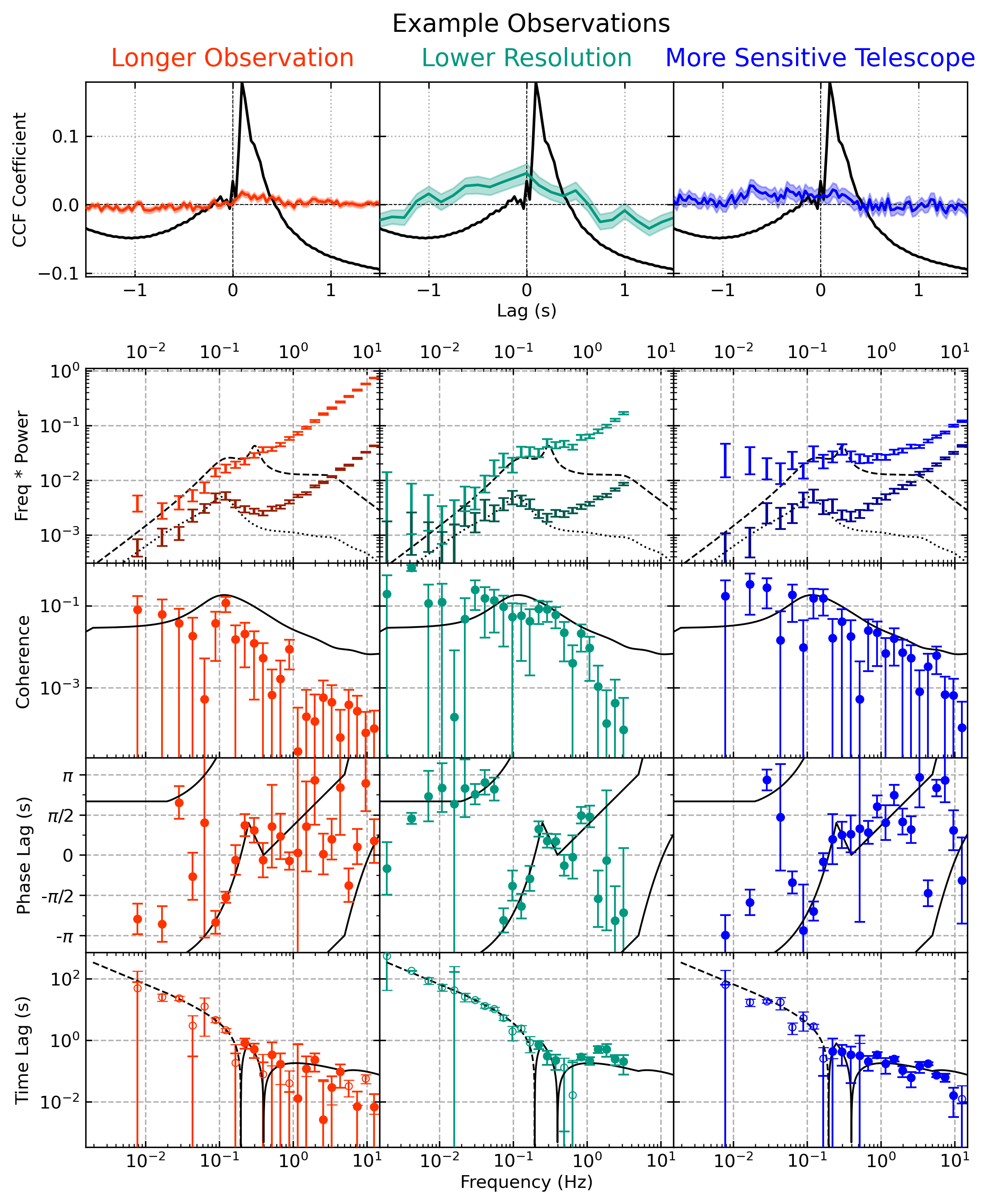}
\caption{Three modified runs of Figure \ref{fig:CorrSimExample_Output}; double the observation length, four times lower time resolution, and a five times more sensitive telescope with half the observing time. Each simulation has been averaged over eight bins. Note how each option has benefits and drawbacks.}
\label{fig:CorrSimExample_Compilation}
\end{figure*}

By altering the parameters, our astronomer runs the program with all these alternative possibilities. Figure \ref{fig:CorrSimExample_Compilation} shows the results.

The resulting simulations show that the longer observation gives a good approximation to the phase lags. Meanwhile, the lower resolution observation gives the highest normalisation in the CCF, and less uncertainty in the coherence values. Finally, the better telescope has a lower white noise floor (i.e. the noise does not dominate the power spectra until higher frequencies than usual; note the difference in power spectra).

These are all important in different ways. For the investigation the 0.1\,s lag offers lower uncertainties around the 0.1-1\,Hz range, the longer observation offers lower uncertainties overall, and the more sensitive telescope has a lower white-noise floor at higher frequencies. Meanwhile, the lower time resolution better replicates the frequency range of the unique feature in the phase lags.

Either way, they now have this additional information with which they can make their decision, which in turn can lead to better data. This ability -- to ask these questions and get informative answers -- is what \texttt{CorrSim} hopes to provide.

\section{Discussion III: Model Testing} \label{sec:Discuss_Model}

So far, we have shown the effect of altering parameters -- both observational and intrinsic -- on resultant lightcurves and their Fourier products, and in particular, the strength of \texttt{CorrSim} in simulating these effects.

Another strength of \texttt{CorrSim} is recreating a previously-carried out observation, and then modifying various Fourier parameters to investigate how they affect other outputs, such as the lightcurves and cross-correlation functions. With this, one can test various different models of a source.

\begin{figure*}
\includegraphics[width=\textwidth]{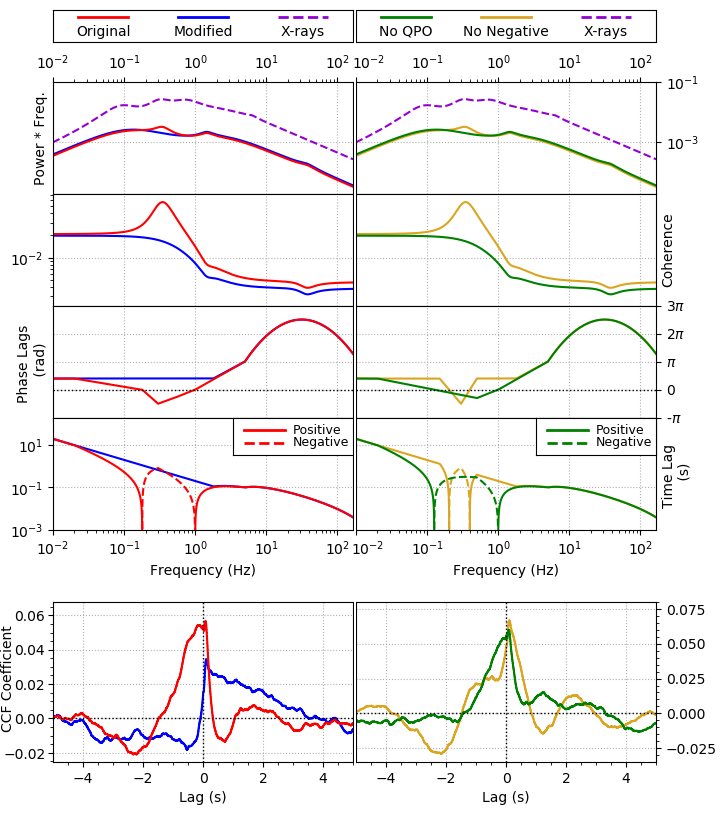}
\caption{Various simulations of MAXI\,J1820+070, as carried out in \citet{Paice_J1820Evolution_2021}. Each column shows two simulations of the $i_{s}$ band (optical) vs X-rays. \textbf{Top:} Input Fourier components. All y-axes are shared. \textbf{Bottom:} CCFs made by converting the Fourier components into lightcurves and then cross-correlating. CCFs were averaged over multiple 10s segments. Note that the y-axes are not shared. \textbf{Left:} The red lines are a close representation of the data. The blue lines are a modification that removes the QPO and the negative lags from the $i_{s}$ band's Fourier components between 0.02--2\,Hz. \textbf{Right:} The green lines are a modification that just removes the QPO, and the gold lines are a modification that just removes the negative lags. Note how the behaviour in the CCF changes between -2 and +3\,s, showing the significance of each component over this range.}
\label{fig:CorrSimEpoch6Combined}
\end{figure*}

This kind of analysis was carried out on MAXI\,J1820+070, which can be seen in Section 4.5 and Appendix A2 of \citet{Paice_J1820Evolution_2021}. The figures showing the simulations have been replicated in Figure \ref{fig:CorrSimEpoch6Combined}. This particular analysis was motivated by a wish to investigate a negative-lag correlation that had appeared in the CCF of one of the epochs; this correlation was not present in any other epoch (most of which had even had a negative-lag \textit{anti-}correlation instead), and was theorised to be connected to a Quasi-Periodic Oscillation (QPO) also present in the source. It was hoped that the Fourier components that were creating this feature could be identified.

This analysis was carried out by first finding the Fourier properties of a real observation of MAXI\,J1820+070, and then modelling them in \texttt{CorrSim}. These properties were then altered to create four variations: the original, unedited model of the source; an edited model with the QPO removed; an edited model with a stretch of negative lags removed (but keeping the QPO); and an edited model with both the QPO and the negative lags removed.

By creating simulated CCFs from each of the models, it was thus determined that the negative-lag correlated component was caused by a mix of both the QPO and the negative lags working constructively. Removing each independently created CCFs which still show a negative-lag correlation, but at a lower significance. Removing both completely removes the component, instead giving a CCF profile with a slight anti-correlation at negative lags and an extended correlation at positive lags, similar to those seen earlier in the source, or even in other sources such as Swift\,J1753.5-0127 \citep{durant_swift_2008}.

From this, the importance of both of these features was concluded: Firstly, that the QPO had a significant effect on the CCF, and such features should be considered in future observations as a possible obfuscating factor; Secondly, that a relatively small extent of negative lags can have a significant effect on the resultant CCF, even without the added coherence of a QPO.

Studies like this can be valuable for future observations. They can allow for testing of alternative Fourier inputs to see how variability changes -- and thus which components are most important. They can also allow for investigating which Fourier inputs and combinations create similar CCFs, thereby testing alternative explanations for already-known features. In either case, these studies have the potential to give valuable insight into the inner processes of XRBs.

\vspace*{0.3cm}
\noindent
    
\section{Conclusions}

First and foremost, we have introduced \texttt{CorrSim}, a tool for simulating two correlated lightcurves based on a variety of parameters, both intrinsic to the source (e.g. mean count rate, power spectra, time lags), and the observation (e.g. observation length and telescope diameter).

Through this program, we have shown how a myriad of different parameters can affect the results of an observation; both parameters intrinsic to the source and caused by our own observations. We have also shown how these effects are not always clear; a higher amount of relative noise is expected to affect the variability and show up in the power spectra, but does not typically lead one to imagine a reduction in the correlation function
, as we see in the supplementary material.

Some of these parameters can affect observations in significant, but sometimes unclear, ways; the effects of a small telescope or a low altitude target are known to be detrimental, but sometimes such parameters cannot be helped and an astronomer may wonder about the quality of the data they could obtain despite this. This program presents a way to answer these questions.

And these answers can have a very practical side; more than just adjusting expectations, one can adjust the parameters of the observation to make sure that any data gained will be of high-enough quality to answer any questions being investigated.

This is the first aim of the program: To provide astronomers with both qualitative and quantitative metrics with which they can assess the quality of the resultant data, and then change the observational setup to optimise them. In this way, the program can be a tool to actively increase the quality of future observations.

We also demonstrated how this program can be used to simulate different properties of a source, showing a practical example that was carried out in \citet{Paice_J1820Evolution_2021}. There, the phase lags were manipulated to remove a range of negative lags, a Quasi-Periodic Oscillation, and then both, in order to see the effect on the correlation function. From these simulations, it was concluded that the presence of a correlation at negative optical lags is probably not an indication of an optical process occurring before an X-ray process, but instead is more likely a periodic X-ray process having an optical response at a lag greater than $\pi$\,radians in phase.

This demonstrates the second aim of this program: to allow astronomers the ability to take models of already-obtained data and change them in order to better understand their makeup, assess their validity, and thus inform them of the components of the system.

Through these abilities, \texttt{CorrSim} can be of significant assistance to astronomers across the timeline of an observation and, it is hoped, will help in increasing the frequency, usefulness, and quality of rapid correlated observations in the future.

\vspace*{0.3cm}
\noindent

\section*{Acknowledgements}
	
We acknowledge support from STFC and a UGC-UKIERI Thematic Partnership. Over the course of this research, JAP was supported by several sources; a University of Southampton Central VC Scholarship, the ERC under the European Union’s Horizon 2020 research and innovation programme (grant agreement No.715051; Spiders), and STFC consolidated grant ST/X001075/1. JAP also thanks D Ashton for spectral timing help, as well as A Stevens and D Huppenkothen for help with the Stingray software. Some simulations were based on observations by GTC/HiPERCAM and NICER; we thank the observers for the HiPERCAM observations, Vik Dhillon and Stuart Littlefair, and we also thank Keith Gendreau, Zaven Arzoumanian, and the rest of the NICER team for their assistance. SMARTNet helped to coordinate those observations. This research made use of software and web tools from the High Energy Astrophysics Science Archive Research Center (HEASARC).

\section*{Data Availability Statement}

All data used within this paper (i.e. all simulated data on which the figures were generated) will be shared on reasonable request to the corresponding author. The program used to generate these data is available at \href{https://gitlab.com/astro_johnapaice/CorrSim}{https://gitlab.com/astro_johnapaice/CorrSim}.


\section*{GLOSSARY} \label{sec:Glossary}

Table \ref{tab:symbol_list} gives a list and definition of all the symbols used in this paper.


\begin{table} 
	\caption{List of Symbols}
	\begin{tabular}{ll}
		\hline
        $A(t)$ & Signal A \\
        $B(t)$ & Signal B \\
        $C(f)$ & Measured Cross-Spectrum \\
        $c(\tau)$ & Correlation Coefficient \\
        $D$ & Normalisation for $L$ \\
        $dT$ & Time Resolution \\
        $F_{rms}$ & Fractional RMS \\
        $f$ & Frequency; dependant variable of $C$, $L$, $y$ \\
        $f_0$ & Midpoint for $L$ \\
        $f_{seg}$ & Number of Frequencies per Segment \\
        $i$ & Imaginary Number \\
        $L(f)$ & Lorentzian \\
        $m$ & Number of Segments \\
        $N$ & Number of Bins \\
        $n$ & Noise (In Power Spectra) \\
        $p$ & Power Spectra \\
        $q$ & Formula shorthand in error on $\gamma^2_I$ \\
        $R_{rms^2}$ & Fractional RMS normalisation \\
        $S$ & Complex Series \\
        $s$ & Signal (Noiseless Power Spectra) \\
        $T$ & Observation Length \\
        $t$ & Time; dependant variable of $A$, $B$, $X$ \\
        $U$ & Amplitude (in Power Spectral calculations) \\
        $V$ & Normalisation (in Power Spectral calculations) \\
        $X(t)$ & Inverse Fourier Transform of $y(f)$ \\
        $y(f)$ & Fourier Transform of $X(t)$ \\
        $\Gamma$ & Full-Width at Half-Maximum for $L$ \\
        $\gamma^2_I(f)$ & Intrinsic Coherence \\
        $\gamma_{Model}$ & Model Coherence \\
        $\Delta T_{samp}$ & Sampling Interval \\
        $\delta$ & Lags \\
        $\delta_{\tau}$ & Time Lags \\
        $\delta_{\phi}$ & Phase Lag \\
        $\tau$ & Lag; dependant variable of $c$ \\
		\hline
	\end{tabular}
	\label{tab:symbol_list}
\end{table}

\bibliography{ref}
\bibliographystyle{mnras}
	

	
	
\appendix
	
\section{Methods} \label{sec:Method}

Analysis of the relationship between two signals can be split into two main themes; those that take place in the time domain (e.g. cross-correlation analysis), and those that take place in the frequency domain (e.g. Fourier analysis). Each have their own benefits and drawbacks.


Cross-correlation analysis gives a coefficient for how two signals are `correlated' based on some lag (measured in time, or bins). Cross-correlation analysis can be useful even with relatively low numbers of bins (compared with Fourier analysis; Figure \ref{fig:CorrSimCompare_ObsLengthNoise} demonstrates how input Correlation models can be reasonably reproduced at small observation lengths, while Fourier models cannot). However, the result of cross-correlation analysis does not trivially translate into quantitative constraints on source properties. For example, multiple interfering processes may `average-out' the resultant coefficient at certain lag ranges, and it can be difficult to separate source features from noise, as well as estimate confidence intervals. Cross-correlation analysis also does not work well when the lags between the two bands are caused by some, potentially variable, periodic process - the resultant values may not show this unique cause.

Cross-frequency analysis, however, ideally solves these two issues; it allows one to determine the time lag as a function of frequency, and the errors on the coherence and the time lags are better quantified \citep[e.g.][]{Vaughan_Nowak_1997}. However, this analysis typically requires a greater number of bins in the lightcurves, and a higher Signal-to-Noise Ratio (S/N). Additionally, if the time-lag is larger than the time period (i.e. the response in the second lightcurve occurs more than one oscillation away from the cause in the first lightcurve), the reported lag will be much shorter, and the true lag obscured; in this case, it is not clear what the full implications are for the measured value and its error. Investigating these questions and determining which analysis is most ideal is one of the motivations behind this work.

Here we will detail the theoretical and mathematical underpinning of both methods.

\subsection{Cross-Correlation Analysis} \label{sec:Method_CCF}

\begin{figure}
	\includegraphics[width=\columnwidth]{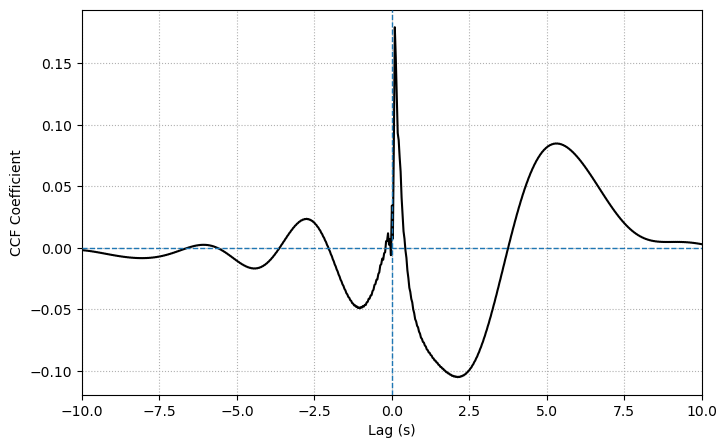}
	\caption{An example CF from cross-correlation analysis, based on data from the black hole X-ray binary MAXI\,J1820+070 \citep{Paice_1820Letter_2019}. Note how an anti-correlation can be seen at a lag of +2\,s, while a correlation can be seen at +5\,s.}
	\label{fig:example_ccf}
\end{figure}

Correlation Functions (CFs) find the relationships between two signals as a function of lag. They produce a `correlation coefficient', between -1 and +1, at each of a range of lag bins; -1 indicating pure anti-correlation, and +1 indicating pure correlation. For example, if there is a sharp rise in Signal A, and then a similar sharp rise in Signal B delayed by 10\,seconds, then a Correlation Function of Signal B vs Signal A will show a positive correlation coefficient at +10\,s. Inversely, if Signal B shows a dip instead of a rise, then the correlation coefficient at +10\,s will be negative instead -- an anti-correlation. An example Correlation Function, with its component signals, is seen in Figure \ref{fig:example_ccf}.
	
\texttt{CorrSim} uses the Cross-Correlation Function (CCF). This particular Correlation Function is described in \citet{VenablesRipley_ModernAppliedStatistics_2002}, and is best suited to simultaneously sampled lightcurves (which we will be using):


\begin{equation}
c(\tau) = \frac{1}{N} \sum_{s=\mathrm{max}(1,-\tau)}^{\mathrm{min}(N-\tau,N)} [A(t+\tau)-\overline{A}][B(t)-\overline{B}]
\label{eqn:correlation}
\end{equation}

where $A(t)$ and $B(t)$ refer to the two signals dependant upon time $t$\footnote{\citet{VenablesRipley_ModernAppliedStatistics_2002} use $X_i$ and $X_j$ for these, and define the coefficient as $c_{ij}(t)$; These have been changed so that symbols are consistent across this paper.}, $\tau$ is the dependant variable (the `lag' in units of bins) where $\tau = 1, ... , N$, and $N$ is the total number of bins in the signals.

The CCF can be modelled mathematically from Fourier components, and is done so by \texttt{CorrSim}. We describe this process in Section \ref{sec:ModelCF}.

There is no straightforward method to evaluate uncertainties or errors in correlation functions. \texttt{CorrSim} uses a method of splitting up the lightcurves into segments, creating a CF from each segment, and then finding the standard error between these individual CFs; this is the method also used by \texttt{crosscorr} in Xspec. However, there are other methods available, though not programmed into \texttt{CorrSim}; Bootstrapping, for instance, is a method of drawing a random selection of points (sometimes with replacement) and then calculating properties from that selection; here, that can take the form of finding the spread of values from a bootstrapped selection of individual CFs. More advanced methods include error analysis carried out by the \texttt{Javelin} code (see \citealt{Zu_Javelin_2011}), or from \citet{MisraBoraDewangan_CCFError_2018} who calculate an analytical formulation for CF errors. 

\subsection{Fourier Analysis}

Fourier analysis, also known as cross-spectral analysis, is founded on the idea that any signal can be constructed by summing together a series of sinusoids of varying phases and amplitudes. Taking two simultaneous lightcurves, we can pair up those component sinusoids which are at the same frequency. One can then determine the coherence and lags between two signals as a function of frequency, rather than having a value at one time lag for all frequencies.

These results can then be used to help disentangle the processes between bands which might be difficult to uncover otherwise. For example, in cross-correlation analysis, strongly-correlated processes could hide weakly-correlated ones; however, if the stronger processes only occur at lower frequencies and the weaker ones at higher frequencies, then Fourier analysis would separate them out. This is especially useful for sources which have processes across the Fourier spectrum -- again, such as X-ray binaries \citep[see][Ch. 2]{Lewin_van_der_Klis_2006}. Additionally, the errors on the coherence and the time lags are better quantified than cross-correlation analysis \citep{Vaughan_Nowak_1997}.

The Fourier transform that is used in our work is the Fast Fourier Transform (FFT) from the SciPy package\footnote{\url{https://docs.scipy.org/doc/}}. There, the equation for a Fourier Transformed series $y$ dependant upon frequency $f$ is:

\begin{equation}
y(f) = \sum_{t=0}^{N-1} e^{-2 \pi i \frac{ft}{N}}X(t)
\label{eqn:fouriertransform}
\end{equation}

along with its inverse,

\begin{equation}
X(t) = \frac{1}{N} \sum_{f=0}^{N-1} e^{2 \pi i \frac{ft}{N}}y(f)
\label{eqn:inversefouriertransform}
\end{equation}

where $X(t)$ is the series to be Fourier transformed, of length $N$, and $i$ is the imaginary number\footnote{In the SciPy documentation, these are referred to as $x(n)$, $N$, and $j$ respectively.}. For Fourier analysis, $N$ should be a power of two; this can be problematic when trying to analyse at the lowest frequencies (which may involve up to half of the data not being used), but for the higher frequencies, the lightcurves can be easily split into smaller segments and averaged.

Much of this research has made use of the Stingray\footnote{\url{https://github.com/StingraySoftware/stingray}} python package \citep{Huppenkothen_Stingray_2019}, which is used in the calculation of Fourier components from the simulated lightcurves.

\subsubsection{Power Spectra} \label{sec:Method_PSD}

A power spectrum shows the amount of variability (`power') at each frequency, which is dependant upon the amplitude of the sin waves at that frequency.

`Power' is an umbrella term, and there are several ways to normalise it. In our work, the normalisation that we have used is the Fractional RMS. This is defined by \citet{vanderKlis_RapidVariability_1997} \citep[see also][]{Miyamoto_VariabilityGX339_1991}, and is given by the formula:

\begin{equation}
R_{rms^2} = \frac{2 \Delta T_{samp}}{\bar{X}^2 N}
\label{eqn:fractionalrms}
\end{equation}

Where $\Delta T_{samp}$ is the sampling interval of the data (i.e. the `exposure time' of the telescope), $\bar{X}$ is the mean rate of the series, and N is the total number of bins in the series. The units of this normalisation are (rms/mean)$^2$\,Hz$^{-1}$. This normalisation is useful, particular for our research, since integrating the power spectrum yields the fractional variance of the data \citep{vaughan_characterizing_2003}.

Power spectra can often by described by a series of Lorentzians, which are distributions given by the form:

\begin{equation}
	\centering
	L(f) = \frac{D}{\pi} \frac{\frac{1}{2}\Gamma}{(f-f_0)^2+(\frac{1}{2}\Gamma)^2} 
\end{equation}

where $D$ is the normalisation (controls its magnitude), $\Gamma$ is the Full Width at Half Maximum (FWHM; controls its width), and $f_0$ is the midpoint ($f_0$ = 0 indicates a zero-centred Lorentzian). For an example of a Power Spectrum being modelled by a series of Lorentzians, see Figure \ref{fig:input_ps_lor}.


\subsubsection{Coherence}

Coherence is the magnitude of the complex-valued cross-spectrum; i.e. how much the amplitudes of the sin waves relate to each other. A higher coherence means the two signals are more correlated at that frequency. This value is constrained between 0 (completely incoherent) and 1 (completely coherent).

The coherence is given by \citet{Vaughan_Nowak_1997}, and mathematically can be given thusly. Assume a Power Spectra $p = \lvert{s}\rvert ^2 + \lvert{n}\rvert^2$, where $\lvert{s}\rvert^2$ is the power of the signal, and $\lvert{n}\rvert^2$ is the noise. The coherence function is then:

\begin{equation}
\gamma_{I}^{2}(f) = \frac{\lvert \langle C(f) \rangle \rvert ^2}{\langle \lvert s_A (f) \rvert ^2 \rangle \langle \lvert s_B (f) \rvert ^2 \rangle}
\label{eqn:coherencefunction1}
\end{equation}

\begin{equation}
\lvert \langle C(f) \rangle \rvert^2 = \lvert \langle s_A^* s_B \rangle + \langle s_A^* n_B \rangle + \langle n_A^* s_B \rangle + \langle n_A^* n_B \rangle \rvert ^2
\label{eqn:coherencefunction2}
\end{equation}

where the subscript $I$ denotes intrinsic coherence between noiseless signals, subscripts A and B refer to the two signals, an asterisk denotes the complex conjugate of the power spectra, and $f$ is frequency.

For XRBs, the variability usually gives high powers with relatively high coherence. Using the error formulation of \citet{Vaughan_Nowak_1997}, the errors in this case can be computed like so:

\begin{dmath}
\gamma^2_I = \frac{ \lvert \langle C \rangle \rvert^2-q^2}{\lvert s_A \rvert ^2  \lvert s_B \rvert^2} \times \left( 1 \pm m^{-1/2} \left[ \frac{2q^4m}{(\lvert \langle C \rangle \rvert ^2-q^2)^2} + \frac{\lvert n_A \rvert ^4}{\lvert s_A \rvert ^4} + \frac{\lvert n_B \rvert ^4}{\lvert s_B \rvert ^4} + \frac{m \delta \gamma^2_I}{\gamma^4_I}  \right]^{1/2} \right)
\label{eqn:coherencefunctionerror}
\end{dmath}

where $m$ is the number of segments averaged over, and

\begin{equation}
q^2 = \frac{ \lvert s_A \rvert ^2 \lvert n_B \rvert ^2 + \lvert n_A \rvert ^2 \lvert s_B \rvert ^2 + \lvert n_A \rvert ^2 \lvert n_B \rvert ^2 }{m}
\label{eqn:coherencefunctionerror2}
\end{equation}

\citet{Vaughan_Nowak_1997} also give an equation for high powers and relatively low coherence; however, this option is not included in \texttt{CorrSim} both for simplicity's sake and because this case does not typically arise in XRBs. This is also true for the case of low powers, combined with the fact that no complete solution exists for confidence values under these conditions.

\subsubsection{Phase and Time Lags} \label{sec:Method_Lags}

The phase lags are the phase angle of the complex-valued cross-spectrum; i.e. the offset between the sin waves at each frequency, as a function of their phase. In units of radians, they lie between $-\pi$ and $\pi$, both of which relate to perfect anti-phase, while a value of 0 relates to perfect phase.

The errors on the phase lags are calculated from the coherence:

\begin{equation}
    \delta_{\phi} = \delta_{\phi} \pm \sqrt{\frac{1-\gamma_{I}^2}{2\gamma_{I}^2f_{seg}m}}
\end{equation}

where $\gamma_I$ is the coherence, $f_{seg}$ is the number of frequencies per segment, and $m$ is the number of segments averaged over \citep{BendatPiersol_RandomDataAnalysis_2000, uttley_multi-wavelength_2014}.

Time lags are similar to phase lags, but are, unsurprisingly, a function of time. They are calculated thusly:

\begin{equation}
\delta_{\tau} = \frac{\delta_{\phi}}{2 \pi f}
\label{eqn:timelags}
\end{equation}

where $\delta_{\phi}$ is the phase lag, and $f$ is the frequency of the bin. Errors on the time lags are calculated similarly. Intrinsic drawbacks of phase lags, and possible solutions to such, are discussed in \ref{sec:CorrSim_PhaseToTime}.

\bsp 
\label{lastpage}
\end{document}